\documentclass[aps,prl,reprint,floatfix]{revtex4-1}

\usepackage[pdftex]{graphicx}
\DeclareGraphicsExtensions{.pdf,.png}

\usepackage{grffile}

\usepackage{amsmath, amsthm, amssymb, amsfonts, amsbsy}
\usepackage{mathtools} 
\usepackage{bm}
\usepackage{mathrsfs}  

\renewcommand*{\v}[1]{\boldsymbol{#1}}
\newcommand\bnabla{\boldsymbol{\nabla}}
\newcommand\bcdot{\boldsymbol{\cdot}}
\renewcommand{\div}{\bnabla\bcdot} 
\newcommand{\m}[1]{\v{#1}}

\renewcommand{\d}{\text{d}} 

\DeclareMathOperator{\arctanh}{arctanh}

\newcommand{\Ca}{\mbox{Ca}}

\usepackage[capitalise]{cleveref}
\crefformat{equation}{Eq. (#2#1#3)} 
\Crefformat{equation}{Equation (#2#1#3)} 
\crefrangeformat{equation}{Eqs. (#3#1#4--#5\crefstripprefix{#1}{#2}#6)} 
\Crefrangeformat{equation}{Equations (#3#1#4--#5\crefstripprefix{#1}{#2}#6)} 
\crefmultiformat{equation}{Eqs. (#2#1#3)}{ and~(#2#1#3)}{, (#2#1#3)}{ and~(#2#1#3)} 
\Crefmultiformat{equation}{Equations (#2#1#3)}{ and~(#2#1#3)}{, (#2#1#3)}{ and~(#2#1#3)} 

\begin{document}

\title{Tractionless Self-Propulsion of Active Drops}

\author{Aurore Loisy}
\author{Jens Eggers}
\author{Tanniemola B. Liverpool}
\affiliation{School of Mathematics, University of Bristol - Bristol BS8 1UG, UK}

\date{\today}

\begin{abstract} 
We report on a new mode of self-propulsion exhibited by compact drops of active liquids on a substrate which, remarkably, is tractionless, i.e., which imparts no mechanical stress locally on the surface. We show, both analytically and by numerical simulation, that the equations of motion for an active nematic drop possess a simple self-propelling solution, with no traction on the solid surface and in which the direction of motion is controlled by the winding of the nematic director field across the drop height. The physics underlying this mode of motion has the same origins as that giving rise to the zero viscosity observed in bacterial suspensions. This topologically protected tractionless self-propusion provides a robust physical mechanism for efficient cell migration in crowded environments like tissues.
\end{abstract}


\maketitle

Self-propulsion, in the absence of load, is required by Newton's laws to be force-free. 
This is ubiquitous in the locomotion of microorganisms where inertia is negligible, from swimming bacteria to crawling cells.  By force-free one means that the \emph{total} force exerted by the environment on the motile entity is zero (and the other way around). This is usually achieved by a gradient of nonzero \emph{local forces} on the boundary, as exemplified by cells crawling on a surface which push at the front and pull at the back.
It would appear impossible to  achieve \emph{traction-free} self-propulsion, that is, with local forces which not only sum up to zero but are identically zero everywhere \footnote{We expand on the notions of force-free and traction-free in Supplemental Material}. However, in 
this Letter, we demonstrate that such self-propulsion is indeed possible for active matter, such as the cytoskeleton of living cells. 
This mode of motion is not possible in objects driven at boundaries but can occur only for those driven in the bulk.

Eukaryotic cells have the ability to move in a variety of complex environments and they do so by adapting their mode of migration to the geometrical and physical properties of their surroundings \cite{Liu2015,Paluch2016}. 
Crawling is a mode of motility well characterized experimentally \cite{Verkhovsky1999,Loisel1999,Yam2007} and captured by a variety of physical models \cite{Kruse2006,Keren2008,Shao2012,Ziebert2012,Tjhung2015,Khoromskaia2015}. 
In crawling, forces are transmitted through local friction, which is provided by the focal adhesions that connect the cytoskeleton to a surface or to the extracellular matrix. However these adhesions are structurally unstable at high strain rates \cite{Schwarz2013}.
This makes crawling ineffective for fast migration in tissues, where cells have to squeeze through tiny gaps.
Motion with zero local force does not have this limitation, with profound implications.

A minimal model to study the physical principles of motility consists of a drop of viscous liquid with orientable components that consume energy (active matter) on a flat rigid surface and confined by surface tension.
A drop of active matter generates internal flows all by itself due to energy input from its components~\cite{Ramaswamy2010,Marchetti2013,Saintillan2013,Prost2015,Julicher2018}, and these can cause the drop to move spontaneously~\cite{Sanchez2012,Hawkins2011,Tjhung2012,Tjhung2015,Giomi2014,Blanch-Mercader2013,Callan-Jones2013,Recho2013,Whitfield2016a,Gao2017}. 
Several studies have shown propulsion of active drops on a surface with a number of related models~\cite{Kruse2006,Shao2012,Ziebert2012,Tjhung2015,Khoromskaia2015}. But those invariably involve local forces (in the form of friction) on the substrate.

\begin{figure}[t]
	\centering
	\includegraphics[width=0.95\linewidth]{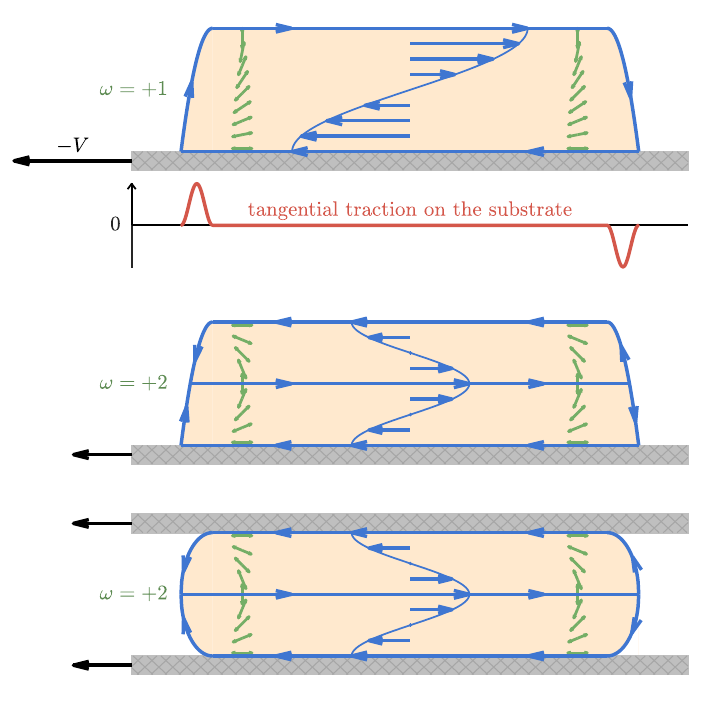}
	\caption{
	Tractionless self-propulsion of an active drop.
	A 
	drop with active stresses $\sigma_{ij}^a = -\alpha n_i n_j$ propels itself (here rightward) at velocity $V$ given by \cref{eq:flat_drop_velocity} without exerting any tangential traction on the substrate (except locally at its edges: this local traction is solely due to surface tension acting to maintain the drop shape).
	The winding number $\omega$, defined as the number of quarter turns of the director $\v{n}$ (green rods) across the drop height, controls the number of fluid circulation cells inside the drop. 
	The internal velocity profile (blue arrows) in the comoving frame of reference is $u_x = - V \cos \left( \omega \pi z/H \right)$, with $z$ being the distance to the substrate, and yields a tank-treading type of motion.
	The solution obtained for $|\omega|=2$ also allows propulsion in a confined geometry.
	\label{fig:scheme_tractionless_tank_treading} 
	}
\end{figure}

In this Letter, we show that an active drop can propel itself without exerting any traction -- defined as the local tangential force per unit area -- anywhere on the substrate, except at its very edge due to surface tension acting to maintain the drop shape (\cref{fig:scheme_tractionless_tank_treading}).
This tractionless self-propulsion is controlled by the global topology of the orientable active units and also allows motion in a confined geometry.
Specifically, we consider a thin drop endowed with active stresses $\sigma_{ij}^a = -\alpha n_i n_j$ ($\alpha \ne 0$ is the signature of activity) induced by active units whose orientations are characterized by a local director $\v{n}$ (a unit vector), and we demonstrate that the equations of motion of this drop on a surface possess a tank-treading solution, which results in a center-of-mass velocity given \emph{exactly} by
\begin{equation}
    V = \frac{\alpha H}{2 \pi \omega \eta},
    \label{eq:flat_drop_velocity}
\end{equation}
where $H$ is a characteristic drop height (whose definition will be made more precise), $\eta$ is the fluid viscosity, and $\omega$ is the (nonzero) winding number defined as the number of quarter turns of the director across the drop height.
The signature of this novel mode of propulsion is the zero tangential traction exerted by the drop on the substrate except at the drop edges due to finite contact angles.

\begin{figure*}
    \flushleft
    (a) \\
    \centering
    \includegraphics[height=3.3cm]{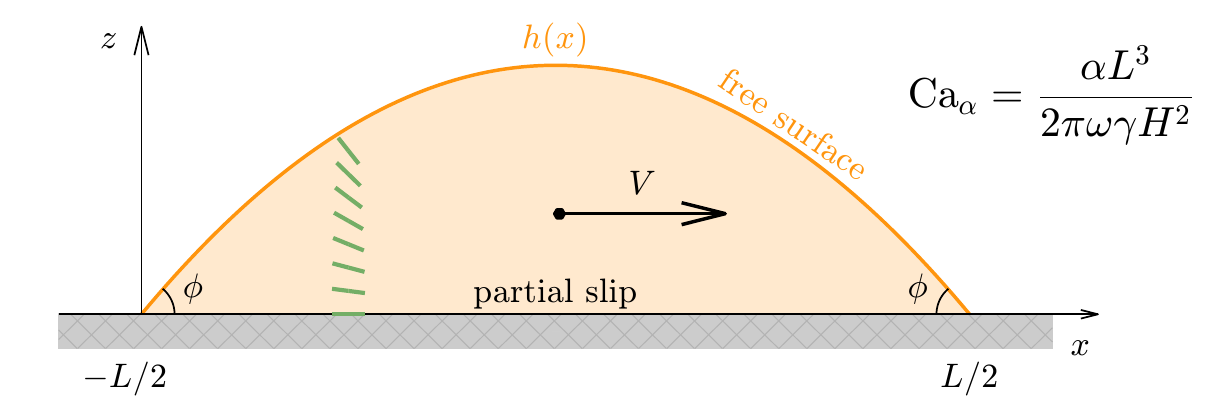} \hspace{0.8cm}
    \includegraphics[height=3.3cm]{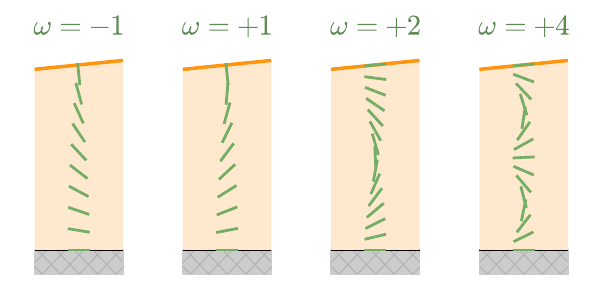} \\[1em]
    \flushleft
    (b) \\
    \centering
    \includegraphics[width=0.99\linewidth]{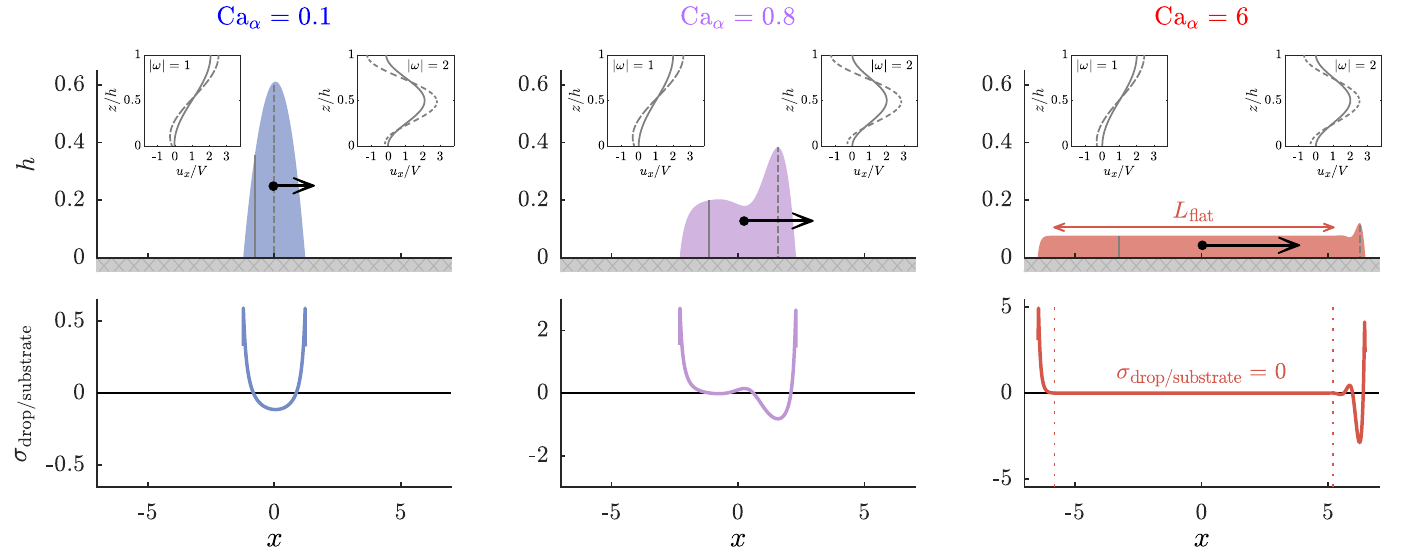} \\
    \flushleft
    (c) \hspace{9.2cm} (d) \\
    \centering
    \includegraphics[height=7.7cm]{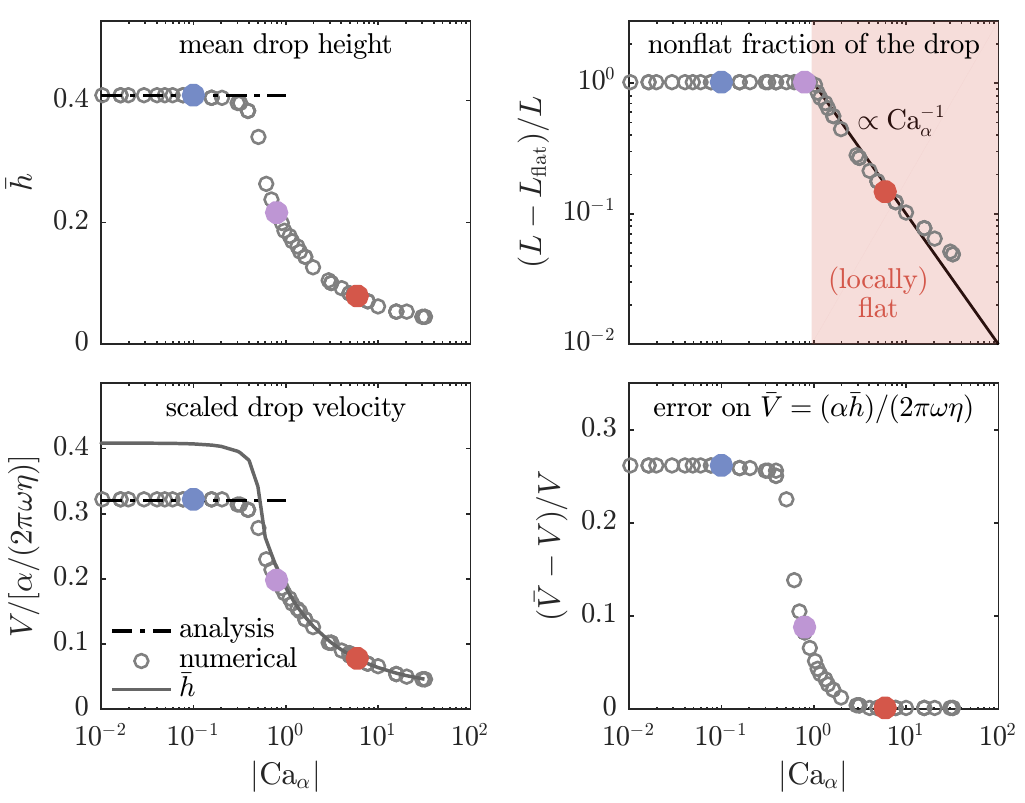} 
    \includegraphics[height=7.7cm]{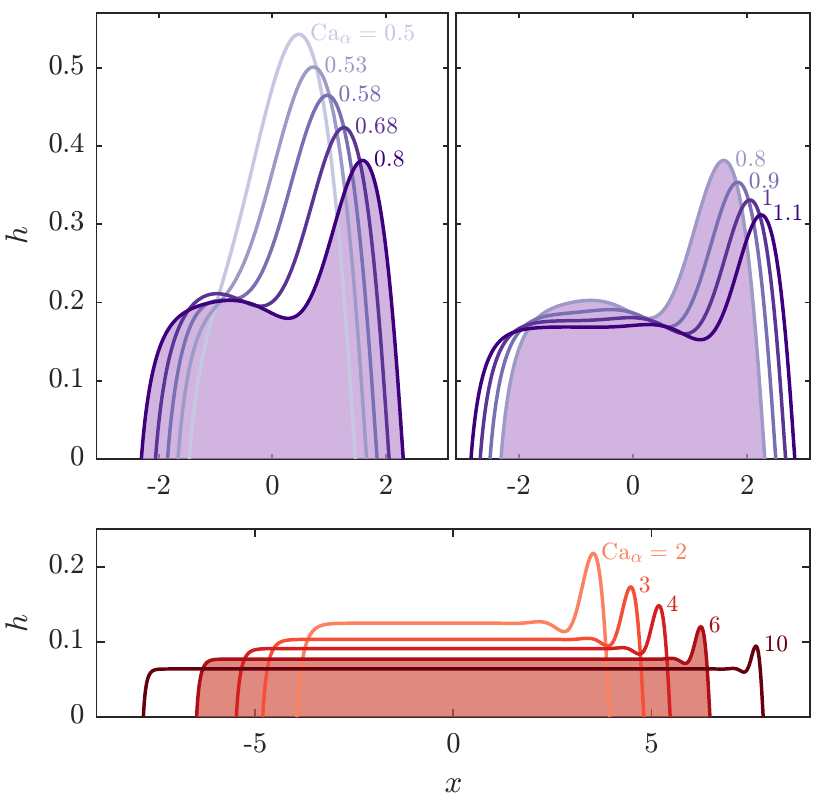}
    \caption{
	(a) Model of a thin active drop with a winded director bounded by a partial slip surface and a free surface. The drop shape and velocity are controlled by $\Ca_\alpha =(\alpha L^3) / ( 2 \pi \omega \gamma H^2)$ where $\omega$ is the winding number.
	(b) Drop shape and tangential traction exerted on the substrate in the three regimes of $\Ca_\alpha$ (insets: internal velocity profiles in selected cross sections for $|\omega|=1$ and $|\omega|=2$).
	(c) Effect of $\Ca_\alpha$ on the drop geometry and velocity: $\bar{h}$ is the mean height of the drop, $(L-L_{\text{flat}})/L$ is the fraction of the drop which is not flat, and $\bar{V}$ is the estimate of the drop velocity based on the exact solution given by \cref{eq:flat_drop_velocity} and using the mean height $\bar{h}$ as an approximation for $H$.
	(d) Dependence of the drop shape on $\Ca_\alpha$ in the transitional regime (top panels) and at high $\Ca_\alpha$ (bottom panel).
	\label{fig:TractionlessTankTreading} 
	}
\end{figure*}

Our model is a 2D drop of active nematic liquid crystal \cite{Ramaswamy2010,Marchetti2013,Prost2015,Julicher2018} moving with (unknown) velocity $V$ on a solid substrate. 
The drop shape is described by its thickness $h(x,t)$ in the normal direction ($z$) as a function of its in-plane position ($x$), as illustrated in \cref{fig:TractionlessTankTreading}a. We seek traveling wave solutions $h=h(x-Vt)$ which must satisfy, from mass conservation,
\begin{equation}
  \int_0^h u_x \; \d z = V h
  \label{eq:unscaled_ODE}
\end{equation}
where $\v{u}$ is the fluid velocity inside the drop, which is assumed to be incompressible ($\partial_i u_i=0$). 

The velocity in \cref{eq:unscaled_ODE} can be obtained by solving the Stokes flow equation \footnote{We have neglected nematic contributions to the stress tensor, which is valid for $H \gg K/\gamma$, with $K$ the Frank elastic constant. We show in Supplemental Material that including nematic stresses has negligible effect on the tractionless self-propelling solution.}
\begin{subequations}
\label{eq:full_Stokes}
\begin{gather}
  \partial_j \sigma_{ij} = 0 \\
  \text{with\ } \sigma_{ij} = - p \delta_{ij} + \eta ( \partial_j u_i + \partial_i u_j ) - \alpha n_i n_j
\end{gather}
\end{subequations}
where $p$ is the pressure and where the director $\v{n}=(\cos \theta, \sin \theta)$ minimizes the nematic free energy in the \emph{strong} elastic limit \footnote{The dynamic equation for the director reduces to $\nabla^2 \theta=0$ provided that $H \ll K/(U \Gamma)$ where $H$ is the characteristic drop height, $U$ is the velocity scale in the direction parallel to the wall, $K$ is the Frank elastic constant, and $\Gamma$ is the rotational viscosity}: $\nabla^2 \theta=0$ with boundary conditions $\theta = 0$ at $z=0$ (anchoring parallel to the substrate) and $\theta = \omega \pi/2 + \arctan(h')$ at $z=h$ (fixed angle $\omega \pi/2$ with respect to the free surface tangent) \footnote{Our results hold for arbitrary anchoring angles, and the strong anchoring condition is relaxed for $h \rightarrow 0$ following the approach of \cite{Cummings2011,Lin2013a}, see Supplemental Material}. In this limit, while the director will affect the flow, the back coupling of the flow on the director is negligible.

At the liquid-solid interface, the interaction of the drop with the rigid substrate is modeled by a partial slip boundary condition: 
\begin{subequations}
\begin{equation}
	u_x = \ell_u \sigma_{xz}/\eta \qquad \text{at $z=0$},
	\label{eq:Navier_slip_BC}
\end{equation}
where $\ell_u$ is a slip length. 
At the gas-liquid interface we use a free surface boundary condition: 
\begin{equation}
	\m{\sigma} \bcdot \v{m} = \gamma \kappa \v{m} \qquad \text{at $z=h$},
	\label{eq:free_surface_BC}
\end{equation}
\end{subequations}
where $\v{m}$ is the unit outward vector normal to the free surface, $\gamma$ is the surface tension, and $\kappa = - \div \v{m}$ is the signed curvature.

Assuming a thin droplet geometry where variations in the $x$-direction are much slower than in the $z$-direction ($\partial_x \ll \partial_z$), as is now standard \cite{Sankararaman2009,Joanny2012,Khoromskaia2015,Kitavtsev2018}, we can derive a leading-order explicit expression of $u_x$ as follows.
It can be shown that if $\omega=0$, the drop is static. Therefore we choose $\omega \neq 0$ and the director orientation reads, at leading order,
\begin{equation}
	\theta =  \frac{\omega \pi}{2} \frac{z}{h}
\end{equation}
and $\omega \in \mathbb{Z}^*$ is effectively a winding number which counts the number of quarter turns of the director across the drop height, as illustrated in \cref{fig:TractionlessTankTreading}a. 

The $z$-component of \cref{eq:full_Stokes} gives, at leading order, $\partial_z p = 0$ and using the normal component of \cref{eq:free_surface_BC} we find that $p= - \gamma h''$.
The $x$-component of \cref{eq:full_Stokes} yields $\partial_z \sigma_{xz} = \partial_x p$. This can be integrated once in $z$, and we find, using the tangential component of \cref{eq:free_surface_BC} and substituting $p$, that
\begin{equation}
	\sigma_{xz} = - \gamma h''' (z-h)
	\label{eq:shear_stress}
\end{equation}
Substituting the definition of $\sigma_{xz}$ into \cref{eq:shear_stress} and integrating once in $z$ with \cref{eq:Navier_slip_BC} yields the parallel velocity
\begin{equation}
  u_x = - \frac{\gamma}{\eta} \left[ \frac{z^2}{2} - (z+\ell_u) h \right] h'''  + \frac{\alpha}{\eta} \frac{(1 - \cos 2 \theta)}{2 \pi \omega} h
  \label{eq:velocity}
\end{equation}
As a last step, we integrate \cref{eq:velocity} over the drop thickness to obtain, using \cref{eq:unscaled_ODE}, a nonlinear third-order ordinary differential equation for the steady-state shape of the drop
\begin{equation}
 \frac{\gamma}{\eta} \left( \frac{h^2}{3}+\ell_u h \right) h''' + \frac{\alpha}{\eta} \frac{1}{2 \pi \omega} h = V
 \label{eq:height_equation_final}
\end{equation}
on the domain $x \in [-L/2,L/2$] with four boundary conditions at the contact lines:
\begin{equation}
 h(\pm L/2) = 0, \qquad h'(\pm L/2) = \mp \phi
 \label{eq:BC}
\end{equation}
where $\phi$ is the prescribed contact angle.
The drop velocity $V$ is an unknown constant determined as part of the solution, and the drop width $L$ is determined by the volume constraint (set here to unity)
\begin{equation}
 \int_{-L/2}^{L/2} h \; \d x = 1.
 \label{eq:volume_constraint}
\end{equation}


Dimensional analysis of \cref{eq:height_equation_final} indicates that the drop shape is controlled by $\Ca_\alpha =(\alpha L^3) / ( 2 \pi \omega \gamma H^2)$, which can be understood as the active equivalent of the capillary number $\Ca_\alpha = (\eta V L^3) / (\gamma H^3)$ with $V$ given by \cref{eq:flat_drop_velocity}.
Since $\Ca_\alpha(\alpha,\omega)=\Ca_\alpha(-\alpha,-\omega)$, changing the direction in which the director winds (from counterclockwise to clockwise) is equivalent to changing the sign of activity (from extensile to contractile). Furthermore, if $\{h(x),V\}$ is a solution for $\Ca_\alpha$ then $\{h(-x),-V\}$ is a solution for $-\Ca_\alpha$: reversing the sign of $\Ca_\alpha$ simply reverses the direction of motion.

Stable solutions of \cref{eq:height_equation_final,eq:BC,eq:volume_constraint} were obtained by solving numerically the time-dependent problem until it reaches steady state for various $\Ca_\alpha$ (see the Supplemental Material for details about numerical methods, in all simulations $\ell_u=0.05$ and $\phi=1$). This allowed us to identify two regimes of low and high $\Ca_\alpha$ separated by a transition region around $\Ca_\alpha \approx 0.8$, as shown in \cref{fig:TractionlessTankTreading}b,c.

For low $\Ca_\alpha$, the drop shape is close to a parabola (the equilibrium solution for a passive drop) and a perturbation analysis gives the solution (shown with dash-dotted lines in \cref{fig:TractionlessTankTreading}c):
\begin{equation}
 V = \frac{\alpha}{2 \pi \omega \eta} \frac{\phi L}{4} \frac{\sqrt{b} - (b-1) \arctanh(1/\sqrt{b})}{\arctanh(1/\sqrt{b})}
 \label{eq:V_anal_pert}
\end{equation}
where $b=1 + 12 \ell_u/(L \phi)$ and $L= \sqrt{6/\phi}$.

At intermediate $\Ca_\alpha$, the drop undergoes a continuous transition to a qualitatively different shape over a narrow range of $\Ca_\alpha$, here for $0.58 < \Ca_\alpha < 1$ (the numerical values of the boundaries of the transition regime depend on $\ell_u$ and $\phi$). First, at $\Ca_\alpha=0.58$, the drop develops a frontal bump ($h(x)$ goes from having one extremum to three extrema, see \cref{fig:TractionlessTankTreading}d, top left panel). Then, at $\Ca_\alpha = 1$, the drop starts flattening at the rear (\cref{fig:TractionlessTankTreading}d, top right panel).

For high $\Ca_\alpha$, the drop spreads and the extent of the flat region grows rapidly with increasing $\Ca_\alpha$ (\cref{fig:TractionlessTankTreading}c, top right panel). Analysis of numerical data indicates that after the transition the nonflat fraction of the drop initially decreases as $\Ca_\alpha^{-1}$. 
The drop velocity can be shown from the governing equation to be given exactly by \cref{eq:flat_drop_velocity}, where $H$ is the local height of the drop where $h'''=0$. In practice, $V$ can be estimated by using the mean drop height as an approximation for $H$ (\cref{fig:TractionlessTankTreading}c, bottom right panel). The error is 5\% at $\Ca_\alpha = 1$ and goes to zero as $\Ca_\alpha \rightarrow \infty$.

The tangential traction exerted by the drop on the substrate is $\sigma_{\text{drop/substrate}} = \sigma_{xz} \rvert_{z=0} = \gamma h h'''$, therefore $\sigma_{\text{drop/substrate}}$ induced by a moving flat drop is identically zero except at the drop edges (\cref{fig:TractionlessTankTreading}b, right panel). It also follows that the drop motion is unaffected by wall slip $\ell_u$. Integrating $\sigma_{\text{drop/substrate}}$ over an edge yields an inward force of magnitude $F_{\text{edge/substrate}} = \gamma \phi^2 / 2$. As seen from the substrate, the drop effectively acts as a contractile force dipole due to finite contact angle $\phi$, independent of the sign and magnitude of activity.

This leads us to offer a very simple sketch of the tractionless self-propulsion of a flat drop in \cref{fig:scheme_tractionless_tank_treading}. The winding of the director generates an active stress in the liquid which must be balanced by the viscous stress such that the total shear stress vanishes. This creates a number of fluid circulation cells inside the drop exactly equal to $|\omega|$.
More precisely, the velocity profile is sinusoidal (compared to parabolic in other modes of motion) and reads in the laboratory frame of reference:
\begin{equation}
 u_x = V \left[ 1 - \cos \left( \frac{\omega \pi z}{H} \right) \right].
\end{equation}
The net flow is not zero and causes the drop to tank tread at a speed $V$ given by \cref{eq:flat_drop_velocity}. Inside the drop, the rate of viscous dissipation $D^\eta=(\eta \pi^2 \omega^2 V^2 L)/(2 H)$ is exactly compensated for by the power input $D^\alpha=-(\alpha^2 H L)/(8 \eta)$ due to activity. The drop speed is maximal for $|\omega|=1$. Higher winding numbers, while less favored energetically, are observed in passive nematics with weak anchoring boundary conditions \cite{Crespo2017} and therefore may still be relevant.

Note that for $|\omega|=1$, $\sigma_{xz}=0$ while the macroscopic strain rate $S=H^{-1} \int_0^H \partial_z u_x \d z = 2 V/H$ is nonzero, meaning that the apparent viscosity of this drop, defined by $\eta_{\text{app}}=\sigma_{xz}/S$, is zero. An analogous mechanism is responsible for the superfluid-like behavior of active suspensions under shear \cite{Lopez2015,Loisy2018b,Guo2018}. 

The solution for $|\omega|=2$ has zero fluid velocity \emph{and} zero shear stress at \emph{both} boundaries, and therefore is also a solution for a drop propelling itself in a channel (bottom panel in \cref{fig:scheme_tractionless_tank_treading}). Traction maps would show a zero traction on the channel walls everywhere except at the drop edges (the direction and magnitude of these localized forces depend on the wettability of the walls). The latter setup is perhaps the easiest to control experimentally: one can imagine confining a drop of bacterial suspension~\cite{Lopez2015,Guo2018} or of microtubule-kinesin mixture~\cite{Sanchez2012} between two surfaces, one used for imaging the traction maps \cite{Tan2003,duRoure2005}  
and the other prepared to ensure appropriate anchoring (through manipulation of the surface chemistry or architecture: see, e.g., \cite{Koumakis2014,Sipos2015,Hasan2013}). 

Unlike other modes of autonomous motion, tractionless tank treading is effective even in the absence of local friction and is not associated with shape anisotropy.
These distinctive features are characteristic of contraction-based amoeboid motility under conditions of weak adhesion and strong confinement, such as those exhibited by leukocyte and human breast cancer cells squeezing through complex 3D extracellular geometries \cite{Lammermann2008,Poincloux2011} or by confined cells migrating in microchannels \cite{Liu2015,Hung2013}. 
Because it does not rely on adhesion, this novel mode of self-propulsion provides an efficient mechanism for fast cell motion in crowded environments. 

Tractionless self-propulsion is driven by the internal fluid flow in the bulk and is not affected by the details of the edges near the contact line, therefore we expect this mechanism to extend directly to 3D. 
It is interesting to make a comparison with driving due to a constant shear stress at the free surface (e.g. Marangoni or wind stresses) \cite{Cazabat1990,Oron1997,Craster2009}, because it gives rise to an identical driving term in the height equation  (\ref{eq:height_equation_final}). Hence we expect the frontal ``bump'' seen in 2D to turn into a fingering instability in 3D. However this similarity is rather superficial as these boundary-driven advancing fronts have very different internal mechanics from our problem: without the internal stress balance and the activity-driven bulk internal flows, they cannot exhibit tractionless self-propulsion.
Finally, the internal flow, caused here by a prescribed winding of the director, could also be induced spontaneously \cite{Voituriez2005}; we leave the analysis of this more mathematically intricate problem for future work.

In conclusion, we have shown that autonomous propulsion, which is always force-free, can also be traction-free. In other words, active materials, by virtue of being bulk driven, can move without imparting any force on their environment. This rather counterintuitive result, derived analytically and confirmed by numerical simulations, is a new example of the astonishing mechanical behavior exhibited by active fluids, like spontaneous flows \cite{Simha2002,Voituriez2005,Sanchez2012,Wu2017}, zero viscosity \cite{Lopez2015,Guo2018} and nonmonotonic flow curves \cite{Loisy2018b}.


\begin{acknowledgments}
We thank Uwe Thiele for helpful discussions and for suggesting the comparison with Marangoni-driven fronts. 
Part of this work was funded by Leverhulme Trust Research Project Grant No. RPG-2016-147.
T.B.L. acknowledges the support of BrisSynBio, a BBSRC/EPSRC Advanced Synthetic Biology Research Centre (Grant No. BB/L01386X/1).
\end{acknowledgments}



%

\end{document}